\documentstyle[epsf]{ioplppt}       % use this for journal style
\begin{document}
\jl{3}
\paper {The generalized localization lengths in one dimensional
    systems with correlated disorder}[The generalized localization
    lengths]
\author{Imre Varga\dag\ddag\ and J\'anos Pipek\dag}
\address
{\dag\ Department of Theoretical Physics, Institute of Physics,
Technical University of Budapest, H-1521 Budapest, Budafoki \'ut 8,
Hungary}
\address
{\ddag\ Condensed Matter Research Group of the
Hungarian Academy of Sciences,
Technical University of Budapest, H-1521 Budapest, Budafoki \'ut 8,
Hungary}
\date{\today}
\begin{abstract}
    The scale invariant properties of wave functions in finite samples
    of one dimensional random systems with correlated disorder are
    analyzed. The random dimer model and its generalizations are
    considered and the wave functions are compared. Generalized
    entropic localization lengths are introduced in order to
    characterize the states and compared with their behavior for
    exponential localization. An acceptable agreement is obtained,
    however, the exponential form seems to be an oversimplification in
    the presence of correlated disorder. According to our analysis in
    the case of the random dimer model and the two new models the
    possibility of power--law localization cannot be ruled out.
\end{abstract}


\section{Introduction}
\label{sec:intro}

    In a previous publication \cite{pv1} we have introduced a new form
    of information length in order to characterize the shape of wave
    functions in finite one dimensional (1d) disordered systems. Using
    that definition we succeeded to show that the states in the 1d
    Anderson model with uncorrelated, onsite disorder do have, apart
    from oscillations, an overall exponential shape. Such an
    exponential decay has been found for practically any strength of
    disorder even in the case when the localization length exceeded
    the size of the system substantially.

    The scaling properties of one--particle states in the presence of
    uncorrelated disorder in 1d and quasi--1d has been studied
    extensively both numerically and analytically \cite{GO4, CGIFM,
    MiFy1, FyMi1}. The similar problem of the more realistic case of
    correlated disorder has been recently considered in \cite{IKT0,
    IKT}. In this paper we wish to present a scale invariant study on
    a wider family of correlated disorder in 1d and at the same time
    show how generalized localization lengths may help to analyze the
    properties of the one--particle eigenstates.

    To be more specific the eigenvalue problem of an electron in a 1d
    disordered potential can be given as
\begin{equation} \label{eq:ham1}
    E\,c_m=\varepsilon_m\,c_m + V_{m,m+1}\,c_{m+1} + V_{m-1,m}\,c_{m-1}
\end{equation}
    where $c_m$ is the amplitude for the electron to be on site $m$
    and $E$ is the energy eigenvalue. In the most simple case studied
    in \cite{pv1} the onsite potentials $\varepsilon_m$ are chosen
    randomly from a box distribution of width $W$ centered around the
    origin, and the off--diagonal hopping integrals are kept constant,
    $V_{m,m+1}=V_{m-1,m}=V_0$. The latter condition enables us to fix
    the unit of the energy scale $V_0=1$. For this model there are
    rigorous results \cite{ishii} affirming complete exponential
    localization for any strength of disorder in {\it infinite}
    systems and in \cite{pv1} we have proved numerically the above
    statement holds in {\it finite} systems, as well.

    The effect of correlated disorder introduced in (\ref{eq:ham1})
    has attracted much attention recently. These correlations maybe
    originated from interactions of electrons with lattice vibrations
    or e.g. as a more realistic representation of disorder
    incorporating the presence of chemical bonding. The first models
    that included such correlations were therefore based on random
    binary models \cite{Masek,KW,JK,ESC,EW,DWP}. These studies
    revealed the possibility that disorder correlations may increase
    and as well as decrease the localization length substantially.

    In a special family of random binary alloys correlation was
    introduced by assigning the same energy level $\varepsilon_A$ or
    $\varepsilon_B$ to pairs of sites. This is called the random
    binary dimer model (RBDM). This model was first introduced in
    \cite{DWP} and it has been shown that under certain conditions
    there are special $E_c$ values where the state is delocalized,
    transparent and the number of extended states around $E_c$ is
    proportional to the square root of the length of a finite sample.
    The localization length diverges at these energies \cite{IKT0,
    Bov} and this is reflected in the conduction properties
    of finite samples \cite{DGK}. In a recent paper \cite{IKT}
    Izrailev {\it et al.} have studied the scaling properties of the
    eigenstates in the RBDM and succeeded to show that the states
    approaching $E_c$ are described similarly as in the case of
    uncorrelated disorder \cite{pv1, CGIFM}.

    There are other similar models formulated in the same spirit as
    the RBDM. A continuous Kronig--Penney type random dimer model
    \cite{SA} exhibits an infinite number of resonances with zero
    reflection constant. The existence of similar special states in a
    quasiperiodic dimer model has also been found in \cite{FGP}.

    Exponential localization, although with an enhanced localization
    length has been seen in another model where the $\varepsilon_m$
    energies are drawn from a box distribution but they are repeated
    for $L$ consecutive sites \cite{SVE}. Here it is possible to vary
    $L$, however, no special energy with complete delocalization is
    present.

    In this paper we present numerical results of two generalized
    versions (A, B) of the RBDM that are related to both the model of
    finite correlation length in \cite{SVE} and the original model of
    Dunlap, Kundu and Phillips \cite{DKP}. Our results are compared to
    the ones obtained by Izrailev {\it et al.} \cite{IKT}.

    In {\sl Model A} the onsite energies are drawn from a box
    distribution and assigned to two consecutive sites at the same
    time: it may be called general random dimer model (GRDM). This
    model is intermediate between the original Anderson model and the
    RBDM, it is in fact the special case of $L=2$ of the model studied
    in \cite{SVE}. As it has been shown \cite{SVE} within such models
    the energy band does not contain any special energies where
    complete delocalization may occur, however, correlations change
    localization in a direction similar to the RBDM.

    {\sl Model B} \cite{DKP} on the other hand contains disorder in
    both the diagonal and the off--diagonal part of (\ref{eq:ham1})
\begin{eqnarray} \label{eq:dkp1}
\varepsilon_m &= {G\over\gamma}V_0(\alpha_{m,m+1}+\alpha_{m,m-1})
        \nonumber\\
    V_{m,m\pm 1}&=
    V_0\sqrt {1+\alpha_{m,m\pm 1}^2-2\alpha_{m,m\pm 1}\cos\delta}
\end{eqnarray}
    where the quantities $\alpha_{m,m\pm 1}$ are chosen from a box
    distribution centered around the origin with width $W\leq 2$. This
    model is obtained considering the coupling of electrons to the
    vibrations of the underlying lattice represented by the random
    variables $\alpha_{m,m\pm 1}$ that introduce a correlated disorder
    in both the onsite and the off--diagonal matrix elements (see
    \cite{DKP} for the details). The correlation is perfect if in
    (\ref{eq:dkp1}) $G=\gamma$. The RBDM can be considered as a
    simplified version of this model. The special energies where
    delocalization occurs are at $E_c=2V_0\cos\delta$. We chose $V_0$
    as the unit of energy here and varied the energy in the vicinity
    of $E_c$ for different values of $\delta $.

    The solution of the Schr\"odinger equation~(\ref{eq:ham1}) using
    appropriate initial conditions $c_o=c_1=1$ is obtained by
    numerically iterating for a system of $N=10^4$ sites and the
    localization properties of the eigenstates are calculated using
    the charge distribution $Q_m=|c_m|^2$. Averaging is performed over
    $M=1000$ samples.

\section{Shape analysis}
\label{sec:shape}

    The shape of the charge distribution $Q_m$ may be characterized
    using the inverse participation number (IPN), $D$, and the
    Shannon--entropy, $H$ \cite{pv1}
\begin{equation} \label{eq:DH}
    D^{-1}=\sum_m\,Q_m^2 \qquad {\rm and}\qquad
    H =-\sum_m\,Q_m\ln Q_m.
\end{equation}
    Both parameters $D$ and $\exp (H)$ give the number of sites
    effectively populated by the state. Therefore a state extending
    over the whole system would have $D=N$ and $H=\ln N$. This means
    that one may introduce two parameters \cite{IKT}
\begin{equation} \label{eq:izr_b12}
    \beta_1={1\over N}\exp (\overline H-H_{ref})
            \qquad {\rm and}\qquad
    \beta_2=\overline D/D_{ref},
\end{equation}
    where normalization with respect to both the system size and the
    case of the absence of disorder has been performed. The latter is
    achieved by evaluating the values $D_{ref}$ and $H_{ref}$ for the
    Bloch--wave solution of (\ref{eq:ham1}) with $\varepsilon_m=0$ and
    $V_{m,m\pm 1}=V_0$ \cite{IKT}. Overbar means average over the
    samples at fixed energy and/or disorder. It is clear that both of
    these quantities change from 1 to 0 as disorder increases from 0
    to $\infty$ therefore in \cite{IKT} they have been used as scaling
    functions.

    Another way of expressing $\beta_1$ and $\beta_2$ ~(\ref{eq:b12})
    can be obtained using our previous definitions \cite{pv1}. There
    we have used a size independent form and pointed out its relevance
    in a scale independent shape analysis of the states. In \cite{pv1}
    we have calculated the spatial filling factor, $q$, and the
    structural entropy, $S$ of individual eigenstates as
\begin{equation} \label{eq:qsstr}
    q=D/N \qquad {\rm and}\qquad S_{str}=H-\ln D.
\end{equation}
    These quantities obey the inequalities: $0<q\leq 1$ and $0\leq
    S_{str}\leq -\ln q$. In the absence of disorder the solution of
    (\ref{eq:ham1}) is a plane wave for which $q^0=2/3$ and
    $S^0_{str}=\ln 3-1$. These quantities can be related to the
    reference values in (\ref{eq:izr_b12}) as $q^0=D_{ref}/N$ and
    $S^0_{str}=H_{ref}-\ln D_{ref}$. Note that $q^0$ and $S^0_{str}$
    are independent of the system size while $D_{ref}$ and $H_{ref}$
    are not. Using $q$ and $S_{str}$ we may rewrite equations
    (\ref{eq:izr_b12}) in the form
\begin{equation} \label{eq:b12}
\beta_1=\overline q\exp (\overline S_{str})/\beta_0
    \qquad {\rm and} \qquad
\beta_2=\overline q/q^0,
\end{equation}
    where $\beta_0=q^0\exp(S_{str}^0)\approx 0.7357$. We have to note
    that the quantities $q^0$, $S_{str}^0$ and $\beta_0$ that appear
    in (\ref{eq:b12}) are obtained naturally from the solution of
    (\ref{eq:ham1}) for uncorrelated disorder which is a plane wave
    modulated by an exponentially decaying envelope represented as a
    product: $c_m\sim\exp (-|m-m_0|/\xi)\sin (km +\delta)$ \cite{pv1}.
    The very same conclusion was drawn using a completely different
    method by Fyodorov and Mirlin \cite{FyMi1} based on results of
    quasi-1d systems and that of strictly 1d \cite{AltPr}.

    The main advantage of this reformulation is the application of the
    shape analysis proposed originally in \cite{pv_orig}. We have
    shown in \cite{pv_orig} and in other publications \cite{vpv, pv}
    that our method is applicable for eigenstates composed as a
    product of several simple forms, e.g. an oscillating planewave
    and an envelope characterized by the scale $\xi$. In \cite{vpv}
    our method unambiguously showed the existence of power--law
    delocalized states near the mobiliy edge of a 1d quasiperiodic
    system, where both the bulk and the tail of the envelope played an
    equal role in the analysis of the wave functions.

    It has been shown in \cite{pv_orig} that parameters $q$ and
    $S_{str}$ can be calculated for ideal charge distributions
    analytically and that the relation $S_{str}(q)$ is directly
    connected to the shape of the distribution. Therefore the
    properties of a large set of numerically obtained wave functions
    are compared to ideal curves in the parameter space $(q,S_{str})$
    especially when some control parameters of the system are varied,
    e.g. system size, strength of disorder or energy. Similar
    relation may follow between $\beta_1$ and $\beta_2$. For example
    in the case of exponential localization we obtain
\begin{equation} \label{eq:b1z}
    \beta_1 (z)={{\exp (z)-1}\over {z\,\exp (z)}}
                \exp\left (1-{z\over {\exp (z)-1}}\right )
\end{equation}
    and
\begin{equation} \label{eq:b2z}
    \beta_2 (z)={2\over z}\left ({{\exp z-1}\over {\exp z+1}}\right )
\end{equation}
    where $z=N/\xi$ with localization length $\xi$. $\beta_1(z)$ and
    $\beta_2(z)$ are monotonous functions of $z$, hence the relation
    $\beta_1\to\beta_2$ exists and is also directly connected to
    the shape of the charge distribution. Note that similar analytic
    $\beta_1 (z)$ and $\beta_2 (z)$ functions can be calculated for
    any other type of form functions, e.g. for power--law decay.

\section{Results and discussion}
\label{sec:res}

    In \cite{IKT} Izrailev {\it et al.} fitted a very simple analytic
    form for the relation $\beta_1\to\beta_2$ in the case of the RBDM
\begin{equation} \label{eq:izr_rel}
    \beta_2={{c\beta_1}\over {1+(c-1)\beta_1}}
\end{equation}
    with $c\approx 0.5488$. We will show that this relation is a good
    approximation, however, it fails to describe the states in both
    the localized and delocalized limits. We have to note that a
    similarly simple scaling relation \cite{CGIFM} for the case of
    uncorrelated disorder has been exhaustively studied in
    \cite{FyMi1} and shown the limitations of it.

\begin{figure}
\epsfxsize=8cm
\centerline{\epsfbox{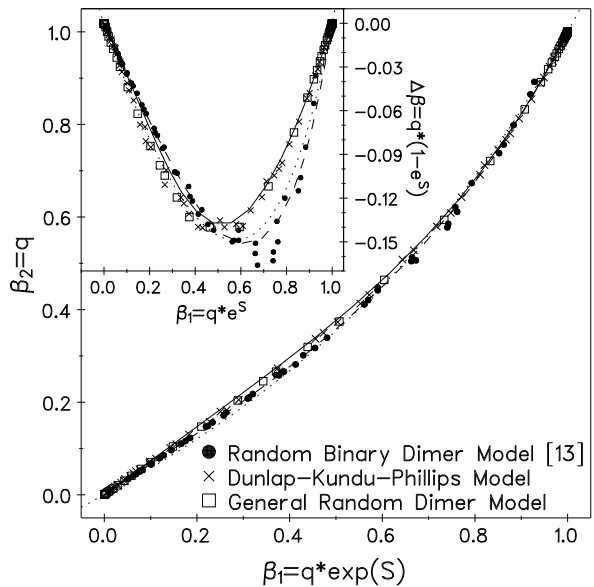}}
\caption{Interrelation $\beta_2 (\beta_1)$ of the generalized
    entropic lengths. Solid symbols are from \protect\cite{IKT}, the
    crosses (DKPM) and open squares (GRDM) are results of the present
    calculation. The continuous curves stand for analytical relations.
    In the inset the difference $\Delta\beta=\beta_2-\beta_1$ is given
    as a function of $\beta_1$. See details in the text.}
\label{fig:b2b1}
\end{figure}

    In terms of $q$ and $S_{str}$ relation (\ref{eq:izr_rel}) reads as
\begin{equation} \label{eq:izr_rel2}
    \tilde S_{str}=-\ln [c+(1-c)\tilde q]
\end{equation}
    where $\tilde S_{str}=\overline S_{str}-S^0_{str}$ and $\tilde
    q=\beta_2=\overline q/q^0$. First of all in the most extended
    limit, $\tilde q\to 1$, $\tilde S_{str}(\tilde q)\approx (1-\tilde
    q)/2$ should hold \cite{pv_orig}. In contrast, according to
    (\ref{eq:izr_rel2}) we get $\tilde S_{str}(\tilde q)\approx (1-c)
    (1-\tilde q)$.

    Secondly we plotted the results of our calculations together with
    the data obtained from \cite{IKT} in \Fref{fig:b2b1}. The
    analytical forms by equations (\ref{eq:b1z},\ref{eq:b2z}) (solid
    curve) and also the empirical relation (\ref{eq:izr_rel})
    (dotted curve) are shown, as well. It is clear that
    (\ref{eq:izr_rel}) is indeed a good approximation, however,
    the tendency is somewhat closer to equations
    (\ref{eq:b1z},\ref{eq:b2z}) that shows exponential localization on
    all length scales. The third relation (dashed curve) is the one
    derived assuming an envelope of the form $c_m\sim m^{-3}$ instead
    of an exponential form. The inset shows the deviation
    $\Delta\beta=\beta_2-\beta_1$ as a function of $\beta_1$. In the
    inset we see again that relation (\ref{eq:izr_rel}) is an
    acceptable fit to the data from \cite{IKT}, however, in this
    figure it is very hard to check its accuracy especially for the
    localized and delocalized limits. In \Fref{fig:b2b1} we have
    plotted our results for {\sl Models A} and {\sl B}, as well. For
    {\sl Model A} we have varied the width of the box distribution
    between $W=10^{-4}V_0$ up to $W=10^4V_0$. In the case of {\sl
    Model B} the parameters $W=1.9V_0$ (here $W$ is limited to $0\leq
    W<2$) and $\delta =0^o$, $30^o$, $60^o$, $90^o$ have been used. We
    can see that, at least in terms of the $\beta_1\to\beta_2$
    relation, both of these models are not very much different from
    the behavior of RBDM.

\begin{figure}
\epsfxsize=8cm
\centerline{\epsfbox{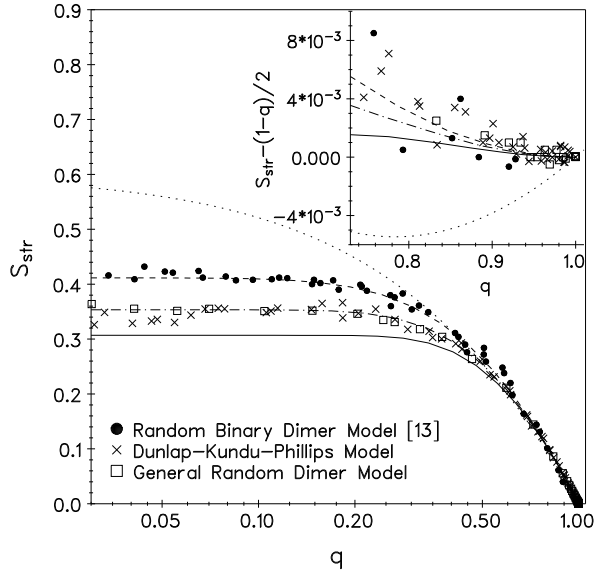}}
\caption{The localization diagram for states in the RBDM, GRDM, and
    the DKPM. The relation (\protect\ref{eq:izr_rel2}) ($\dotted $)
    clearly deviates from the numerical results for $q\to 0$. The
    symbols are like in \protect\Fref{fig:b2b1}. In the inset
    the deviation of $S_{str}$ is given from its expected universal
    form $(1-q)/2$ when $q\to 1$. The dotted line is wrong in this
    limit, as well.}
\label{fig:sq}
\end{figure}

    In order to investigate the similarities and differences between
    the RBDM and the models studied here ({\sl Model A}, the GRDM and
    {\sl Model B}, the DKPM) it is even more apparent to plot
    $S_{str}$ as a function of $\ln q$. In \Fref{fig:sq} the
    localized region $q\to 0$ is clearly not described by the relation
    (\ref{eq:izr_rel2}) presented with a dotted line. It is also true
    that the states in neither models show clear exponential
    localization in the ideal form depicted by the continuous solid
    curve as it is already suggested in the previous paragraph.
    However, the states in {\sl Models A} and {\sl B} studied here are
    closer to it. This means that in the strong localization limit,
    $q\to 0$, the states in the RBDM have definitely more complex
    structure than that of {\sl Models A} and {\sl B}. It is
    interesting to note that similar deviation has been obtained for
    the case of weak uncorrelated disorder in \cite{AltPr}. However,
    in contrast to \cite{AltPr} our results are nonperturbative.
    Furthermore we have to stress that in the present work disorder
    correlations play an important role yielding the above mentioned
    deviations from conventional exponential localization.

    In \Fref{fig:sq} we have also plotted the $S_{str}(q)$
    relation for power--law localization with different exponents. We
    observe that the RBDM is well described by an overall shape:
    $c_m\sim m^{-3}$ (dashed line) while the GRDM and the DKPM are
    better described with $c_m\sim m^{-6}$ (dashed--dotted line).
    Apparently this is in contradiction with analytical expressions of
    the Lyapunov exponent (inverse localization length) which goes as
    $\gamma(E)\sim (E-E_c)^2$ around the special energies $E_c$
    \cite{IKT0, Bov}, however, $\gamma$ should vanish for the case of
    power--law localization \cite{vpv, pv}. A possible resolution to
    this problem is already outlined in \Sref{sec:shape}, e.g. an
    exponential decay superimposed by some kind of rapidly varying
    substructure can easily provide a shift from the curve
    corresponding to the exponential decay to power--law decay as it
    is seen in \Fref{fig:sq}. According to \Fref{fig:sq} the relation
    (\ref{eq:izr_rel2}) is a wrong approximation especially for the
    strong localization limit. Moreover, in the inset of \Fref{fig:sq}
    we see deviations for the delocalized limit, $q\to 1$, as well.

\section{Conclusions}
\label{sec:conc}

    We have performed a shape analysis of wave functions obtained in
    several one dimensional random models with correlated disorder. We
    have introduced a new definition for the generalized localization
    lengths based on the inverse participation number and the
    Shannon--entropy. We have applied the shape analysis introduced in
    \cite{pv_orig}.

    It has been shown that the localization properties of the states
    in the RBDM are described by (\ref{eq:izr_rel}) only
    approximately. On the other hand equation (\ref{eq:izr_rel2})
    shows a wrong behavior for $q\to 1$. Large deviations are obtained
    in the localized limit $\beta_2\to 0$. We have compared the data
    from \cite{IKT} with our simulations on the GRDM ({\sl Model A})
    and the DKPM ({\sl Model B}). The data show a clear deviation from
    simple exponential localization: the average localization
    properties of the states for the RBDM {\it resemble} that of a
    power--law shape $c_m\sim m^{-3}$ and in the case of {\sl Models
    A} and {\sl B} that of a power--law shape $c_m\sim m^{-6}$.

\ack
    One of the authors (I.V.) is grateful for F. Izrailev for
    providing the data of their calculations. Financial support from
    Orsz\'agos Tudom\'anyos Kutat\'asi Alap (OTKA), Grant Nos.
    T014413/1994, T021228/1996, T024136/1997 and F024135/1997 are
    gratefully acknowledged.

\end{document}